\documentstyle[epsf]{mn}
\newcommand{\noi}{\noindent}

\title[Eccentric stellar discs]
{Eccentric stellar discs with strong density cusps and separable potentials}
\author[M. A. Jalali and A. R. Rafiee]
        {M. A. Jalali \thanks{E-mail: jalali@iasbs.ac.ir}
        \& A. R. Rafiee \\
        Institute for Advanced Studies in Basic Sciences, 
        P.O. Box 45195-159, Gava Zang, Zanjan, IRAN}

\begin{document}
\label{firstpage}
\maketitle

\begin{abstract} 
We introduce a class of eccentric discs with ``strong" density 
cusps whose potentials are of St\"ackel form in elliptic coordinates. 
Our models exhibit some striking features: sufficiently close to the 
location of the cusp, the potential and surface density distribution 
diverge as $\Phi \propto r^{-1}$ and $\Sigma \propto r^{-2}$, respectively. 
As we move outward from the centre, the model takes a non-axisymmetric, 
lopsided structure. In the limit, when $r$ tends to infinity, the 
isocontours of $\Phi$ and $\Sigma$ become spherically symmetric. It is 
shown that the configuration space is occupied by three families of 
regular orbits: {\it eccentric butterfly}, {\it aligned loop} and 
{\it horseshoe} orbits. These orbits are properly aligned with the 
surface density distribution and can be used to construct 
self-consistent equilibrium states.
\end{abstract}


\section{INTRODUCTION}
High resolution observations based on {\it Hubble Space Telescope} photometry
of nearby galaxies have increased our understanding of the central regions 
of elliptical and spiral galaxies. It was found that in most galaxies 
density diverges toward the centre in a power-law cusp. In the presence
of a cusp, regular box orbits are destroyed and replaced by chaotic orbits
(Gerhard \& Binney 1985). Through a fast mixing phenomenon, stochastic orbits
cause the orbital structure to become axisymmetric at least near the centre
(Merritt \& Valluri 1996). These results are confirmed by the findings
of Zhao et al. (1999, hereafter Z99). Their study reveals that highly 
non-axisymmetric, scale-free mass models can not be constructed 
self-consistently. Among the models studied for self-consistency, 
one can refer to the integrable, cuspy models of Sridhar \& Touma (1997, 
hereafter ST97). Without a nuclear black hole (BH), centrophobic 
bananas are the only family of orbits presenting in ST97 discs. 
Although such orbits elongate in the same direction as density profile, 
the orbital angular momentum takes a local minimum somewhere rather 
than the major axis where the surface density has a maximum. This is 
the main obstacle for building self-consistent equilibria by regular 
bananas (Syer \& Zhao 1998; Z99). A similar situation occurs for 
anti-aligned tube and high resonance orbits for which one could not 
be able to fit the curvatures of orbits and surface density 
distribution near the major axis (Z99). According to the results 
of Miralda-Escud\'e \& Schwarzschild (1989), it is only possible 
to construct self-consistent models by certain families of fish orbits. 

The orbital structure of stellar systems is
enhanced by central BHs in a different manner.
Although nuclear BHs destroy box orbits, they
enforce some degree of regularity in both centred
and eccentric discs (Sridhar \& Touma 1999,
hereafter ST99). In systems with analytical cores
and central BHs, a family of long-axis tube orbits
can help the host galaxy to maintain its
non-axisymmetric structure within the BH sphere
of influence (Jalali 1999). 

In this paper, we present a class of non-scale-free, 
lopsided discs, which display a collection of properties 
expected in self-consistent non-axisymmetric cuspy systems. 
Our models are of St\"ackel form in elliptic coordinates
(e.g., Binney \& Tremaine 1987) for which the Hamilton-Jacobi 
equation separates and stellar orbits are regular. In central regions 
where the effect of the cusp dominates, the potential functions of 
our distributed mass models are proportional to $r^{-1}$ 
as $r \rightarrow 0$. So, we attain an axisymmetric structure near 
the centre which is consistent with the predicted nature of density 
cusps. The slope of potential function changes sign as we 
depart from the centre and our model galaxies considerably become 
non-axisymmetric. Non-axisymmetric structure is supported by
a family of eccentric loop orbits, which are {\it aligned} with 
the lopsidedness. Our potential functions have a local minimum 
around of which a family of {\it eccentric butterfly} orbits 
emerges. Close to the centre, loop orbits break down and give 
birth to a new family of orbits, {\it horseshoe} orbits. Stars 
moving in horseshoes lose their kinetic energy as they approach 
to the centre and contribute a large amount of mass to form a cusp. 
Our models can be applied to the study of dynamics in systems 
with double nucleus such as M31 (Tremaine 1995, hereafter T95) 
and NGC4486B (Lauer et al. 1996). 

\section{THE MODEL}
Consider the Hamiltonian function
\begin{equation}
{\cal H} = \frac 12 (p_x^2+p_y^2) + \Phi (x,y), \label{1}
\end{equation}
\noindent which is described in cartesian
coordinates, $(x,y)$. The variables $p_x$ and
$p_y$ denote the momenta conjugate to $x$ and
$y$, respectively. $\Phi$ is the potential due
to the self-gravity of the disc. Let us express
${\cal H}$ in elliptic coordinates, $(u,v)$,
through the following transformations
\begin{eqnarray}
x&=&a (1+\cosh u \cos v), \label{2} \\
y&=&a \sinh u \sin v, \label{3} \\
u &\geq& 0,~~0\le v \le 2 \pi, \nonumber
\end{eqnarray}
where $a$ is constant and $2a$ is the distance
between the foci of confocal ellipses and
hyperbolas defined by the curves of constant
$u$ and $v$, respectively. In the new
coordinates, the Hamiltonian function becomes
\begin{equation}
{\cal H} = \frac 1{2a^2(\sinh ^2 u + \sin ^2 v)}
(p_u^2+p_v^2) + \Phi (u,v), \label{4}
\end{equation}
with $p_u$ and $p_v$ being the new canonical
momenta. We think of those potentials which
take St\"ackel form in elliptic coordinates.
The most general potential of St\"ackel form is
\begin{equation}
\Phi (u,v)= \frac {F(u)+G(v)}
{2a^2(\sinh ^2 u + \sin ^2 v)}, \label{5}
\end{equation}
\noi where $F$ and $G$ are arbitrary functions of their arguments. 
By this assumption, the Hamilton-Jacobi equation separates and 
results in the second integral of motion, $I_2$. We get
\begin{equation}
I_2=p_u^2 - 2a^2 E \sinh ^2 u + F(u),  \label{6}
\end{equation}
\noi or equivalently
\begin{equation}
-I_2=p_v^2 - 2a^2 E \sin ^2 v + G(v),  \label{7}
\end{equation}
\noi where $E$ is the total energy of the system,
$E \equiv {\cal H}$. 

We now introduce a class of potentials with
\begin{eqnarray}
F(u) &=& C (\cosh u)^{\gamma}, \nonumber \\
G(v) &=& -C \cos v |\cos v|^{\gamma-1}, \label {8}
\end{eqnarray}
where $C>0$ and $\gamma$ are constant parameters. One can readily 
verify that
\begin{equation}
\cosh u = \frac 1{2a} (r+s),~~
\cos v = \frac 1{2a} (r-s), \label{9}
\end{equation}
\noi where
\begin{equation}
r^2 = x^2+y^2,~~s^2 = (x-2a)^2+y^2. \label{10}
\end{equation}
\noi We substitute from (\ref{10}) into (\ref{8}) and express $\Phi$ 
in the $(x,y)$ coordinates:
\begin{eqnarray}
\Phi &=& K \frac {(r+s)^{\gamma}-(r-s)|r-s|^{\gamma-1}}
{2r s}, \label{11} \\
K &=& C(2a)^{-\gamma}. \nonumber 
\end{eqnarray}
\noi The surface density distribution, associated with $\Phi$, is 
determined as (see Binney \& Tremaine 1987):
\begin{equation}
\Sigma (x',y')=\frac {1}{4 \pi ^2 G}
\int \int \frac {(\nabla ^2 \Phi) dxdy}
{\sqrt {(x'-x)^2+(y'-y)^2}}. \label{12}
\end{equation}
\noi We examine the characteristics of the potential 
and surface density functions for small and large radii. 
Very close to the centre, we have $r \ll s$ that simplifies 
(\ref{11}) as follows
\begin{equation}
\Phi=\frac {Ks^{\gamma -1}}2 \frac {(1+\frac rs)^{\gamma}+
(1-\frac rs)^{\gamma}}{r}. \label{13} 
\end{equation}
\noi One can expand $(1+\frac rs)^\gamma$ and 
$(1-\frac rs)^\gamma$ in terms of $r/s$ to obtain
\begin{equation}
\Phi=\frac {Ks^{\gamma -1}}r \left [ 1+ \sum_{n=1}^{\infty} 
\frac {\Gamma (\gamma+1)}{(2n)!\Gamma (\gamma -2n+1)} 
\left ( \frac rs \right )^{2n} \right ], \label{14} 
\end{equation}
\noi where $\Gamma$ is the well known Gamma function. 
As $r$ tends to zero, $s$ is approximated by $2a$ and
$r/s \rightarrow 0$. Therefore, Equation (\ref{14}) reads
\begin{equation}
\Phi \approx \frac {K(2a)^{\gamma -1}}r. \label{15} 
\end{equation}
\noi Dimensional considerations show that the surface
density $\Sigma$ will approximately be proportional 
to $r^{-2}$. Thus, sufficiently close to the centre, 
we obtain a strong density cusp with spherical symmetry. 
When $r$ tends to infinity, the potential $\Phi$ is 
approximated as
\begin{equation}
\Phi \approx K 2^{\gamma-1} r^{\gamma -2}. \label{16} 
\end{equation}
\noi So, we find out 
\begin{equation}
\Sigma \propto r^{\gamma-3}. \label{17} 
\end{equation}
\noi We have to select those values of $\gamma$ for which 
the surface density distribution is plausible and orbits 
are bounded. According to (\ref{17}), the surface density 
decays outward ($r \rightarrow \infty$) for $\gamma <3$. 
Moreover, Equation (\ref{16}) shows that orbits will be 
escaping if $\gamma \le 2$. To verify this, consider the 
force exerted on a star, which is equal to $-\nabla \Phi$.
This force will always be directed outward for $\gamma \le 2$ 
and results in escaping motions. Therefore, we are confined 
to $2<\gamma<3$. 

\begin{figure}
\centerline{\hbox{\epsfxsize=3.2in\epsfbox{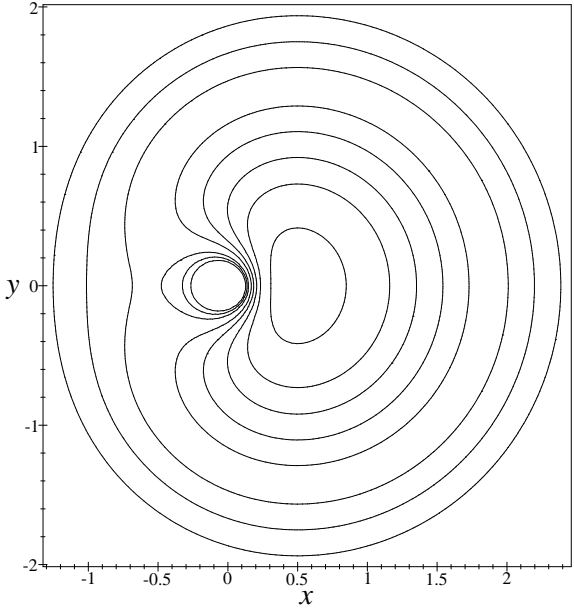}}}
\centerline{(a)}
\centerline{\hbox{\epsfxsize=3.2in\epsfbox{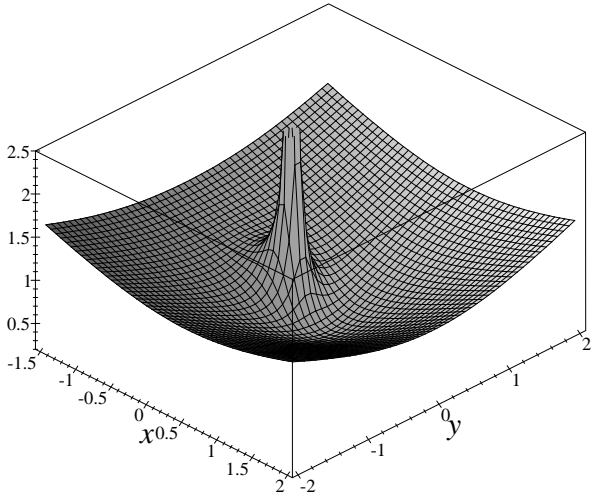}}}
\centerline{(b)}
\caption[Figure 1]{As a demonstrating example, we have determined
the potential $\Phi$ for $\gamma=2.8$, $a=0.5$ and $K=0.2$. Figures 
(a) and (b) show the isocontours and a three dimensional view of
$\Phi$, respectively.} 
\end{figure}

We have used Equations (\ref{11}) 
and (\ref{12}) to compute $\Phi$ (Figure 1) and $\Sigma$ 
(Figure 2) for $\gamma=2.8$. Due to the complexity of 
$\nabla ^2 \Phi$, we have utilized a numerical scheme 
to evaluate the double integral of (\ref{12}). The potential 
and surface density functions are symmetric with respect to 
the $x$-axis and are cuspy at $(x=0,y=0)$. The potential 
$\Phi$ has a local minimum at $(x=a,y=0)$ that plays an 
important role in the evolution of orbits. This minimum point 
has no image in the plane of the surface density isocontours. 
The surface density monotonically decreases outward from the 
centre. As it is evident from Figure 2, a non-axisymmetric, 
lopsided structure is present at moderate distances from the 
centre. 

\begin{figure}
\centerline{\hbox{\epsfxsize=3.2in\epsfbox{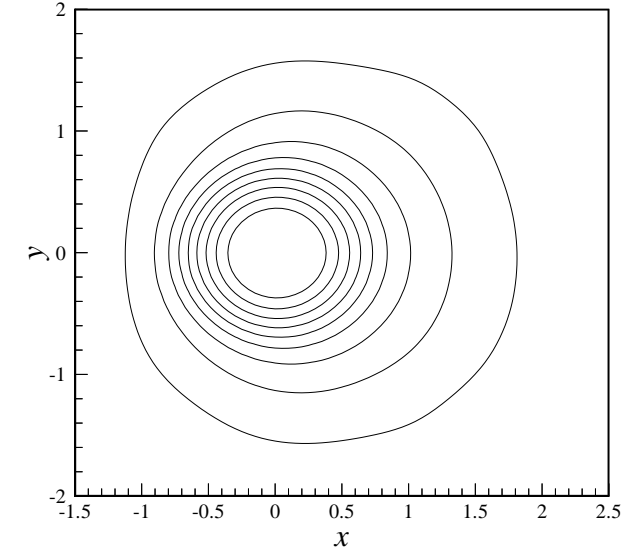}}}
\caption[Figure 2]{The surface density isocontours corresponding 
to the potential function of Figure 1.} 
\end{figure}

\section{ORBITS}
To this end, we classify orbit families.
Having the two isolating integrals $E$ and $I_2$, one can find
the possible regions of motion by employing the positiveness
of $p_u^2$ and $p_v^2$ in (\ref{6}) and (\ref{7}).
We define the following functions:
\begin{eqnarray}
f(u)&=& -2 a^2 E \sinh ^2 u + F(u), \label{18} \\
g(v)&=& -2 a^2 E \sin ^2 v + G(v), \label{19}
\end{eqnarray}
\noi where $F(u)$ and $G(v)$ are given as (\ref{8}).
By virtue of $p_u^2 \geq 0$ and $p_v^2 \geq 0$ one can write
\begin{eqnarray}
I_2-f(u) &\geq& 0, \label{20} \\
-I_2-g(v) &\geq& 0. \label{21}
\end{eqnarray}
\noi Due to the nature of $\Phi$, no motion exists for negative 
energies. Hence, $E$ can only take positive values, $E>0$. 
Our classification is based on the 
behavior of $f(u)$ and $g(v)$. The most general form of $f(u)$ 
is attained for $\gamma C<4a^2E$. In such a circumstance, 
$f(u)$ takes a local maximum at $u=0$, $f_{\rm M}=f(0)=C$, 
and a global minimum at $u=u_{\rm m}$, $f_{\rm m}=f(u_{\rm m})$,  
where 
\begin{equation}
\cosh u_{\rm m}=\left ( \frac {4a^2E}{C\gamma} \right )^
{\frac {1}{\gamma-2}}, \label{22}
\end{equation}
\noi and
\begin{equation}
f_{\rm m}=-2a^2E \sinh ^2 u_{\rm m} + 
C (\cosh u_{\rm m})^{\gamma}. \label{23} 
\end{equation}
\noi According to (\ref{20}) we obtain 
\begin{equation}
I_2 \ge f_{\rm m}. \label{24}
\end{equation}
On the other hand, $g(v)$ has a global maximum at 
$v=\pi$, $g_{\rm M}=g(\pi)=C$, and two global minima 
at $v=\pi/2$ and $v=3\pi/2$, 
$g_{\rm m}$=$g(\pi/2)$=$g(3\pi/2)$=$-2a^2E$.
Therefore, Inequality (\ref{21}) implies
\begin{equation}
I_2 \le 2a^2E. \label{25} 
\end{equation}
By combining (\ref{24}) and (\ref{25}) one achieves
\begin{equation}
f_{\rm m} \le I_2 \le 2a^2E. \label{26} 
\end{equation}
\noi It should be noted that $2a^2E>C$. This is because 
of $2< \gamma < 3$. $f_{\rm m}$ and in consequence $I_2$, 
can take both positive and negative values. Depending on 
the value of $I_2$, three general types of orbits are 
generated:

(i) {\it Eccentric Butterflies}. For $C<I_2<2a^2E$, 
the allowed values for $u$ and $v$ are
\begin{equation}
u \le u_0,~v_{b,1} \le v \le v_{b,2},
~v_{b,3} \le v \le v_{b,4}, \label{27}
\end{equation}
\noi where $u_0$ and $v_{b,i}$ ($i=1,2,3,4$) are the
roots of $f(u)=I_2$ and $g(v)=-I_2$, respectively.
As Figure 3a shows, the horizontal line that indicates
the level of $I_2$, intersects the graph of $f(u)$ at
one point, which specifies the value of $u_0$. The
line corresponding to the level of $-I_2$ intersects
$g(v)$ at four points that give the values of
$v_{b,i}$s (Figure 3b). In this case the motion
takes place in a region bounded by the coordinate
curves $u=u_0$ and $v=v_{b,i}$. The orbits fill
the shaded region of Figure 4a. These are butterfly
orbits (de Zeeuw 1985) displaced from the centre.
We call them eccentric butterfly orbits.

\begin{figure}
\centerline{\hbox{\epsfxsize=1.7in\epsfbox{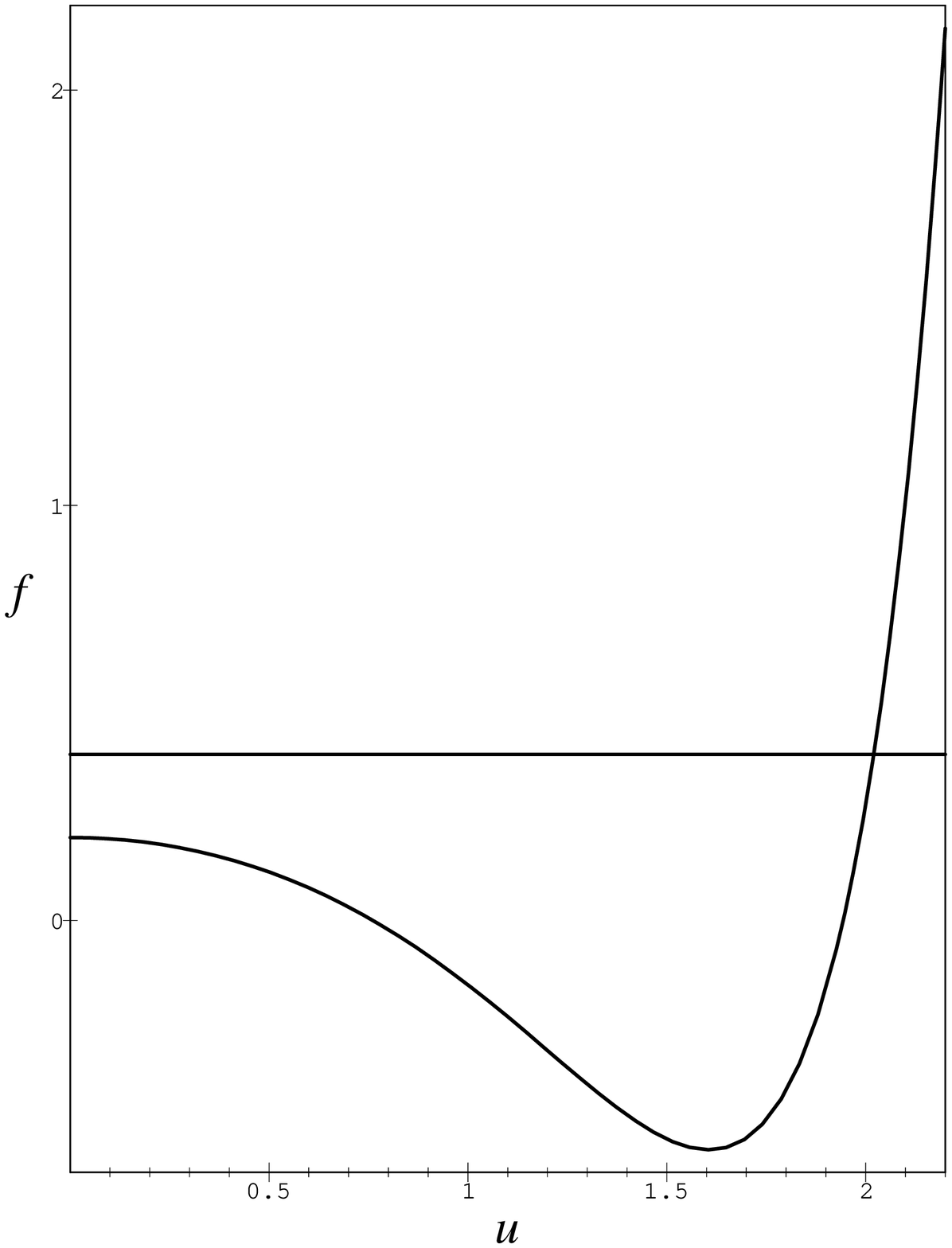}
                  \epsfxsize=1.7in\epsfbox{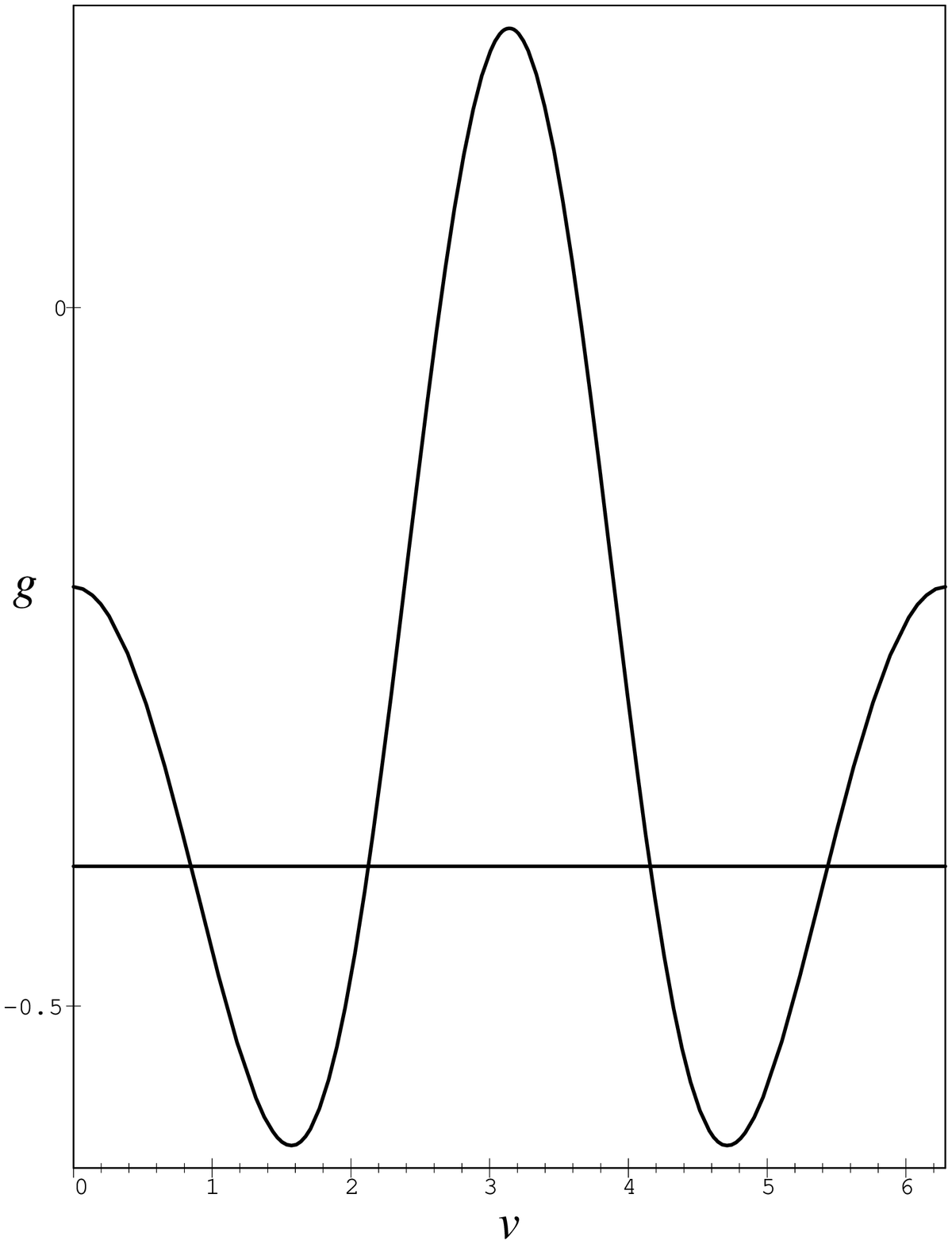}}}
\centerline{\hspace*{0.8in}$(a)$\hfill$(b)$\hspace{0.7in}}
\centerline{\hbox{\epsfxsize=1.7in\epsfbox{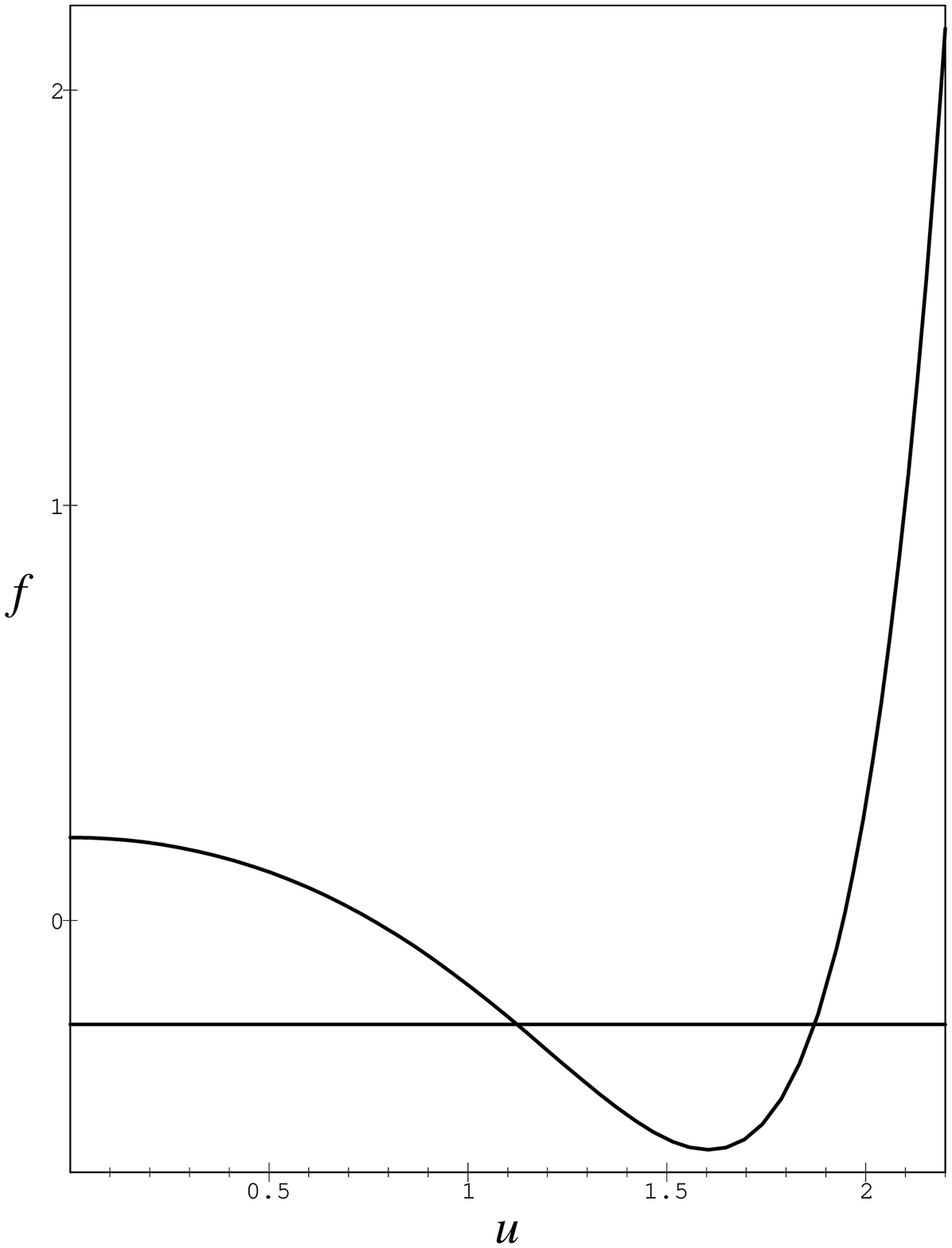}
                  \epsfxsize=1.7in\epsfbox{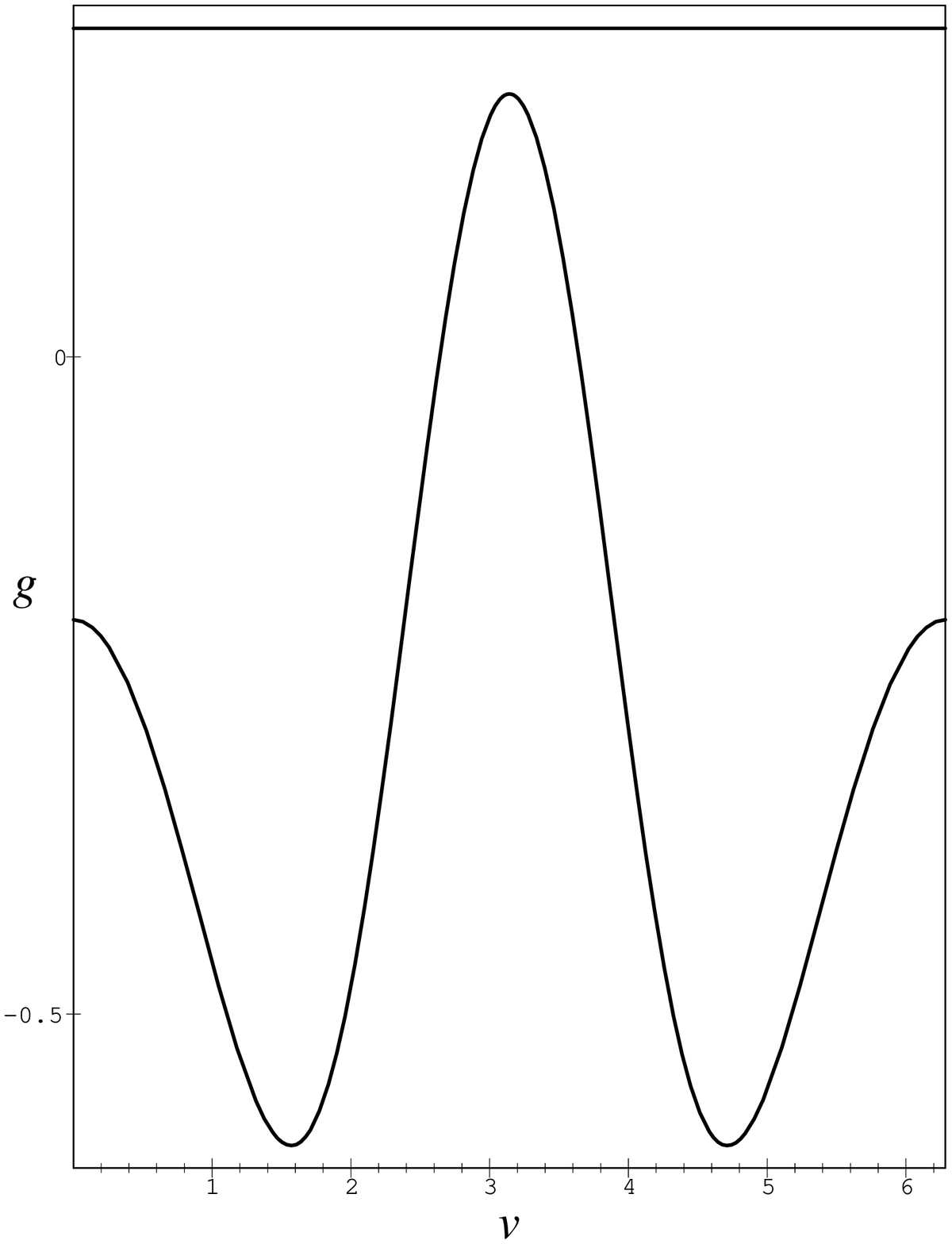}}}
\centerline{\hspace*{0.8in}$(c)$\hfill$(d)$\hspace{0.7in}}
\centerline{\hbox{\epsfxsize=1.7in\epsfbox{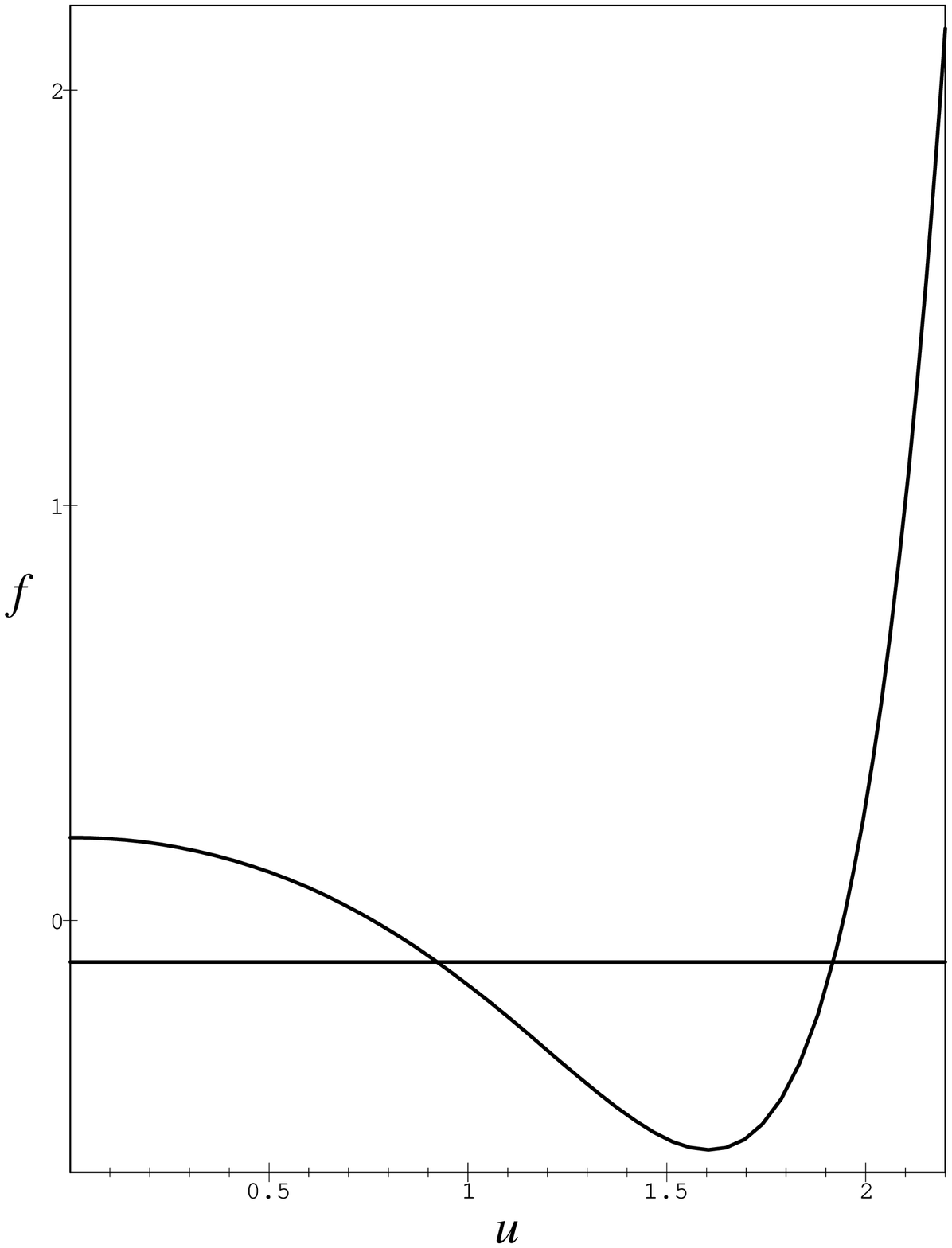}
                  \epsfxsize=1.7in\epsfbox{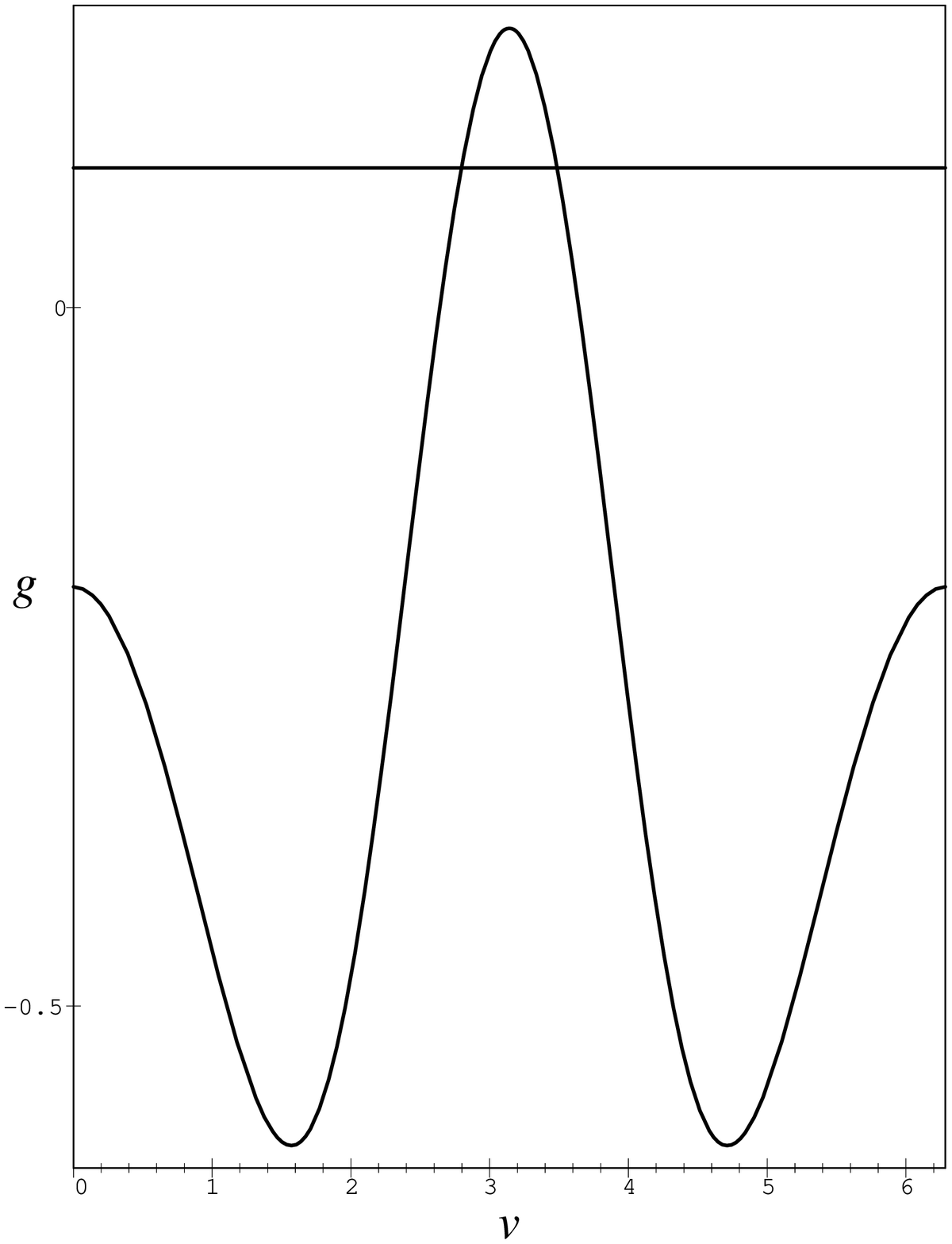}}}
\centerline{\hspace*{0.8in}$(e)$\hfill$(f)$\hspace{0.7in}}
\caption[Figure 3]{The graphs of $f(u)$ and $g(v)$ for
$\gamma=2.8$, $C=0.2$, $E=1.2$ and $a=0.5$. The horizontal
lines indicate the levels of $I_2$ and $-I_2$ in the graphs
of $f(u)$ and $g(v)$, respectively.
(a) $I_2=0.4$ (b) $-I_2=-0.4$ (c) $I_2=-0.25$
(d) $-I_2=0.25$ (e) $I_2=-0.1$ (f) $-I_2=0.1$.} 
\end{figure}

(ii) {\it Aligned Loops}. We now let $I_2$ be negative so
that $f_{\rm m}<I_2<-C$. In this case the equation $f(u)=I_2$
has two roots, $u_{l,1}$ and $u_{l,2}$, which can be identified 
by the intersections of $f(u)$ and the level line of $I_2$ (see
Figure 3c). The equation $g(v)=-I_2$ has no real roots and
Inequality (\ref{21}) is always satisfied (Figure 3d). 
The allowed ranges of $u$ and $v$ will be
\begin{equation}
u_{l,1} \le u \le u_{l,2},~~0 \le v \le 2 \pi. \label{28}
\end{equation}
\noi The orbits fill a tubular region as shown in Figure 4b.
These orbits are bound to the curves of $u=u_{l,1}$ 
and $u=u_{l,2}$ and elongate in the same direction as 
lopsidedness. Following ST99, they are called aligned loops.

(iii) {\it Horseshoes}. For $-C<I_2<C$, we have a
different story. In this case, both of the
equations $f(u)=I_2$ and $g(v)=-I_2$ have
two roots. We denote these roots by $u=u_{h,i}$
and $v=v_{h,i}$ ($i=1,2$). In other words, the
level lines of $\pm I_2$ intersect the graphs of
$f(u)$ and $g(v)$ at two points as shown in
Figures 3e and 3f. The orbits fill the shaded
region of Figure 4c, which looks like a horseshoe.
We call these horseshoe orbits. The orbital
angular momentum of stars moving in horseshoes
($G=xp_y-yp_x$) flips sign when stars arrive
at one of the coordinate curves $v=v_{h,1}$ or
$v=v_{h,2}$.

\begin{figure}
\centerline{\hbox{\epsfxsize=1.7in\epsfbox{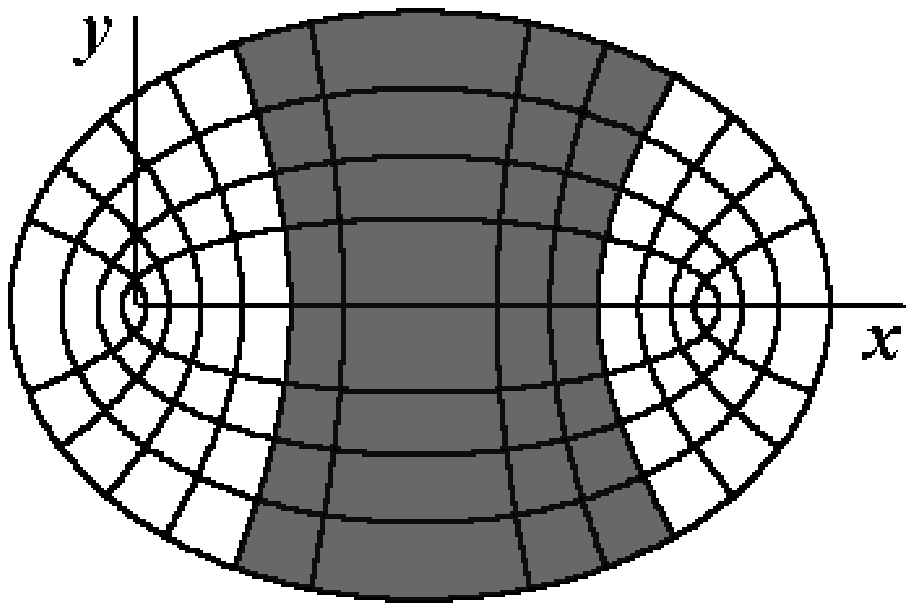}
                  \epsfxsize=1.7in\epsfbox{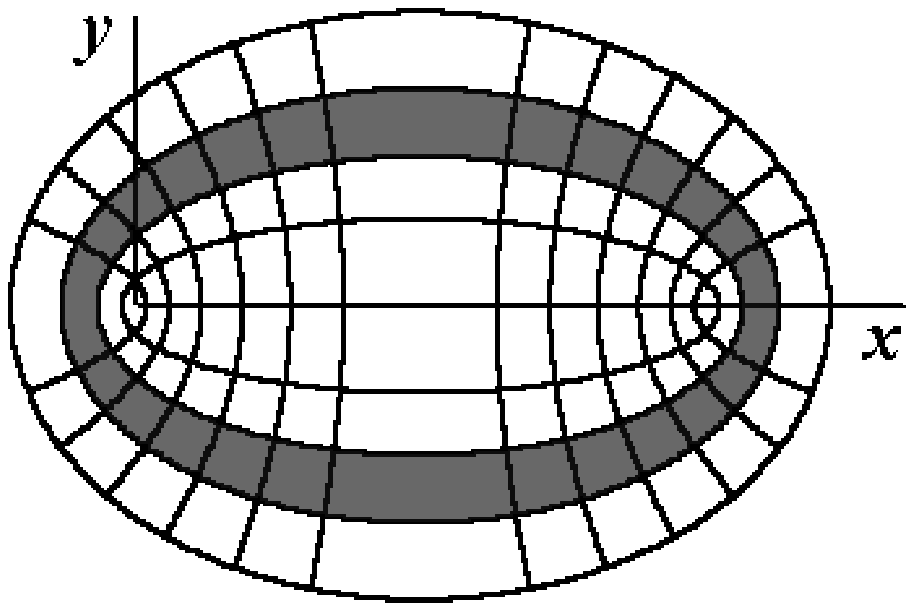}}}
\centerline{\hspace*{0.75in}$(a)$\hfill$(b)$\hspace{0.75in}}
\vspace{0.3cm}
\centerline{\hbox{\epsfxsize=1.7in\epsfbox{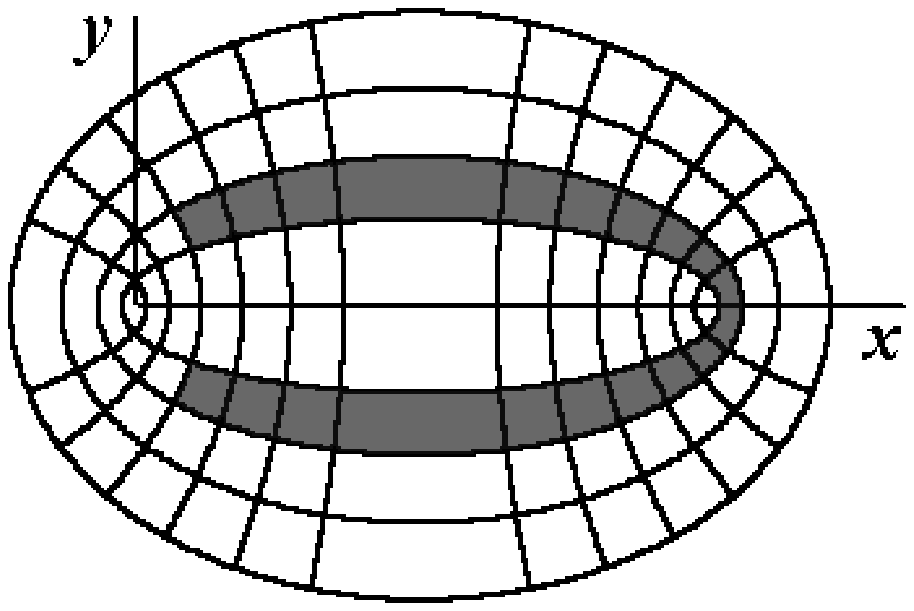}
                  \epsfxsize=1.7in\epsfbox{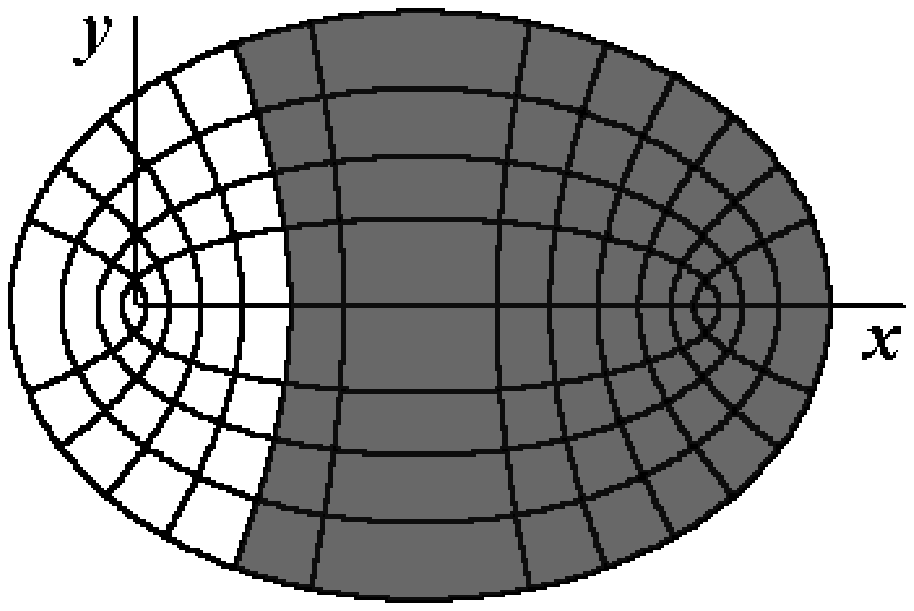}}}
\centerline{\hspace*{0.75in}$(c)$\hfill$(d)$\hspace{0.75in}}
\caption[Figure 4]{The possible families of orbits:
(a) an eccentric butterfly orbit (b) an aligned loop orbit 
(c) a horseshoe orbit (d) a lens orbit associated with $I_2=C$.} 
\end{figure}

For $\gamma C > 4a^2E$, $f(u)$ is a monotonically increasing 
function of $u$ 
and eccentric butterflies are the only existing family of orbits.
There are three transitional cases corresponding to $I_2=C$, 
$I_2=2a^2E$ and $I_2=f_{\rm m}$. For $I_2=C$, eccentric butterflies 
extend to a {\it lens} orbit as shown in Figure 4d. For $I_2=2a^2E$, 
stars undergo a rectilinear motion on the line $x=a$ with the 
amplitude of $\pm a \sinh u_0$ in the $y$-direction. 
For $I_2=f_{\rm m}$, loop orbits are squeezed to an elliptical 
orbit defined by $u=u_{\rm m}$.

\section{DISCUSSIONS}
In this work we explore a credible model based on the 
self-gravity of stellar discs to explain how an eccentric 
disc, with strong density cusp, can be in equilibrium.
Our mass models exhibit most features of eccentric stellar 
systems, especially, double nucleus ones such as M31 and
NGC4486B. 

All of the orbits of our model discs are non-chaotic.
Below, we clarify how the existing families of orbits help
the eccentric disc to maintain the assumed structure.

The force exerted on a star is equal to
$-\nabla \Phi$. The motion under the influence
of this force can be tracked on the {\it potential hill} 
of Figure 1b. This helps us to better imagine the
motion trajectories.

As Figure 1b shows, the potential function is concave.
A test particle released from distant regions with
$x>0$ and ``small" initial velocity, slides down on
the potential hill and moves toward the local minimum
at ($x=a,y=0$). After passing through the neighborhood
of this point (there are some trajectories that exactly
visit the minimum point), the test particle climbs on
the potential hill until its potential energy becomes
maximum. Then, the particle begins to slip down again. 
This process is repeated and the trajectory of the 
particle fills an eccentric butterfly orbit. Stars 
moving in eccentric butterflies form a local group in
the vicinity of ($x=a,y=0$). The accumulation of stars 
around this local minimum of $\Phi$ can create {\it a 
second nucleus} like P2 in M31 (see T95). 
The predicted second nucleus will approximately be 
located at the ``centre" of loop orbits while the 
brighter nucleus (P1) is at the location of the cusp. 

Aligned loop orbits occur when the orbital angular
momentum is high enough to prevent the test particle
to slide down on the potential hill. The boundaries
of loop orbits are defined by the ellipses
$u=u_{l,1}$ and $u=u_{l,2}$. The central cusp is
located at one of the foci of these ellipses.
Aligned loops have the same orientation as the
surface density isocontours (compare Figures 2 and
4b). Thus, according to the results of Z99, it is 
possible to construct a self-consistent model using 
aligned loop orbits.

Similarly, we can describe the behavior of
horseshoe orbits. Stars that start their 
motion sufficiently close to the centre, 
are repelled from the centre because the force
vector is not directed inward in this region.
As they move outward, their orbits are bent 
and cross the $x$-axis with non-zero angular 
momentum. These stars considerably lose their 
kinetic energy as they approach the centre 
(this is equivalent to their climb on the cuspy 
region of the potential hill). Meanwhile, 
the orbital angular momentum takes a minimum 
and switches sign somewhere on the boundary of 
horseshoe orbit. This boundary is defined by 
$v=v_{h,1}$ (or $v=v_{h,2}$) and can be chosen 
arbitrarily close to the centre. These stars 
spend much time near the centre and deposit a 
large amount of mass, which generates a cusp. 
Therefore, horseshoe orbits can be used to 
construct a self-consistent strong cusp. The 
method of Z99 is no longer applicable to horseshoes
because such orbits don't cross the long axis 
(here the $x$-axis) near the centre. In fact, 
horseshoe orbits are an especial class of boxlets 
that appropriately bend toward the centre.  
The lack of such a property in banana orbits causes 
the ST97 discs to be non-self-consistent.

In the case of M31 and NGC4486B, if we suppose that 
loop and high-energy butterfly orbits control the 
overall shape of outer regions, horseshoe orbits
together with low-energy butterflies (small-amplitude 
liberations around the local minimum of $\Phi$) can support 
the existence and stability of a double nucleus. The 
parameter $a$ will indicate the distance between P1 and P2.

There remains an important question: what does happen
to a star just at the centre? The centre of the model,
where the cusp has been located, is inherently unstable.
With a small disturbance, stars located at ($x=0,y=0$)
are repelled from the centre. But, the time that they
spend near the centre will be much longer than that
of distant regions when they move in horseshoes. We
remark that the stars of central regions live in
horseshoe orbits. Although one can place a point
mass (black hole) at the centre without altering
the St\"ackel nature of the potential, such a point
mass will not remain in equilibrium and leaves the
centre. Based on the results of this paper, we 
conjecture that there may not be any mass 
concentration just at the centre of cuspy galaxies.
However, a very dense region exists {\it arbitrarily}
close to the centre! This may be an explanation of
{\it dark} objects at the centre of cuspy galaxies.
The centre of our model galaxies is unreachable.
Our next goal is to apply the method of Schwarzschild
(1979,1993) for the investigation of self-consistency.

%\section{ACKNOWLEDGMENTS}
%Discussions with S. Sridhar were thoughtful.
%This work was supported by IASBS.

\end{document}